\documentclass[intlimits,twoside,a4paper]{article}

\usepackage{amsmath,amssymb}
\usepackage{graphicx}
\usepackage{color}

\usepackage[T2A]{fontenc}
\usepackage[cp1251]{inputenc}

\usepackage[eqsecnum]{cmpj2}

\issue{2014}{17}{3}{33007}
\doinumber{10.5488/CMP.17.33007}

\title[Directional depletion interactions in shaped colloidal particles]
{Directional depletion interactions in shaped particles}
\author[A. Scala, P.G. De Sanctis Lucentini]{A. Scala\refaddr{label1},
        P.G. De Sanctis Lucentini\refaddr{label2}}
\addresses{
\addr{label1} ISC-CNR Physics Department, University ``La Sapienza'', Piazzale Moro 5, 00185 Roma, Italy
\addr{label2} Gubkin Russian State University of Oil and Gas, 65 Leninsky prospekt, Moscow, Russia}

\date{Received April 30, 2014, in final form May 29, 2014}
\authorcopyright{A. Scala, P.G. De Sanctis Lucentini, 2014}

\begin{document}

\maketitle

\begin{abstract}

Entropic forces in colloidal suspensions and in polymer-colloid systems are of long-standing and continuing interest. Experiments show how entropic forces can be used to control the self-assembly of colloidal particles.
Significant advances in colloidal synthesis made in the past two decades have enabled the preparation of high quality nano-particles with well-controlled sizes, shapes, and compositions, indicating that such particles can be utilized as ``artificial atoms'' to build new materials.
To elucidate the effects of the shape of particles upon the magnitude of entropic interaction, we analyse the entropic interactions of two cut-spheres. We show that the solvent induces a strong directional depletion attraction among flat faces of the cut-spheres. Such an effect highlights the possibility of using the shape of particles to control directionality and strength of interaction.
\keywords entropic interactions, self-assembly, Monte Carlo
\pacs 05.20.Jj, 05.10.Ln, 47.57.-s, 65.80.-g, 82.70.Dd
\end{abstract}


\section{Introduction}

{Small particles S being added to a suspension of bigger particles B, induce a ``depletion'' attraction among the bigger particle that arises from an unbalanced osmotic pressure due to the exclusion of small particles from the region between big particles.} Small particles are often referred to as the ``depletant''. The strength of this interaction can be easily explained {by modelling big and small particles by hard-spheres of radii $R$ and $r$; simple Asakura-Oosawa geometric arguments} \cite{AsakuraOosawa54,Vrij76} {show that depletion interaction is attractive and is controlled by the concentration of small particles and the depletant-to-solute ratio $R/r$, while the range of attraction depends on the radius $r$ of small particles.} A classical experimental example is the case of adding non-adsorbing polymers (with radius of gyration $r$) to a stable colloidal solution of silica spheres (with colloids of size $R$) \cite{Yodh01}; at such scales, interactions can be considered to appear only due to steric factors. While the original Asakura-Oosawa model considered only hard-core interactions~--- hence depletion forces have necessarily got an entropic origin~--- in principle, there can be additional thermodynamic contributions from the non-sharp inter-particle repulsion potentials, attractive terms and an explicit contribution of the solvent.

Depletion interactions induce an attractive force among particles that can be tuned both in the strength and in the range; hence, it is possible to induce~--- often via phase transitions  mechanisms~--- a variety of micro-structures such as gels, complex liquids or exotic crystals with the final aim of achieving mechanical and rheological properties suitable for particular applications.

To understand the mechanism of depletion interaction, let us consider a system of two big particles in a bath of depletant particles and assume that only hard interactions are at play. At low depletant concentrations, the free energy of a system of hard particles will be as follows:
\begin{equation}
F=-k_\textrm{B}T\rho V_\textrm{free}\,,
\end{equation}
where $\rho$ is the number density of the depletant particles and $V_\textrm{free}$ is the volume available to depletant; hence, the presence of big particles increases the free energy of the system by reducing the free volume. In particular, if we model the depletant as hard particles of diameter $\sigma$, each big particle will ``occupy'' an excluded volume $V_\textrm{excl}$ that extends {up to a distance $r=\sigma/2$} from its surface. Nevertheless, when big particles come close together (their surfaces are at a distance less than $\sigma$), their excluded volumes overlap and decrease the free energy by a factor proportional to the volume $\Omega$ of the overlap region: all this results in an effective attractive interaction of magnitude $\sim k_\textrm{B}T\rho \Omega$ and range $\sim\sigma$.
It is, therefore, possible to tune the range and the strength of depletion interaction by varying the size $\sigma$ and the packing fraction $\phi=\pi\rho\sigma^{3}/6$ of the depletant.

It is nowadays possible to engineer shaped building blocks at nanometer and micrometer scales. New pioneering techniques permit unprecedented synthesis and fabrication of shaped nano-particles and colloids; shape only can induce exotic phases \cite{Ellison06} as it is  well known from liquid crystal science \cite{Allen93}. The final goal of producing self-assembling materials requires the capability of controlling and designing the interactions among the building blocks. While depletion interactions in colloidal and nano systems are of long-standing and continuing interest, entropic forces among anisotropic particles started to be systematically investigated only in the recent decades \cite{Eisenriegler05,Li03,Lu2003,Li05}. In this paper, we intend to further develop the investigation of entropic forces among anisotropic particles.

\section{Model and simulations}

\begin{figure}[htb]
\centerline{
\includegraphics[width=0.25\textwidth]{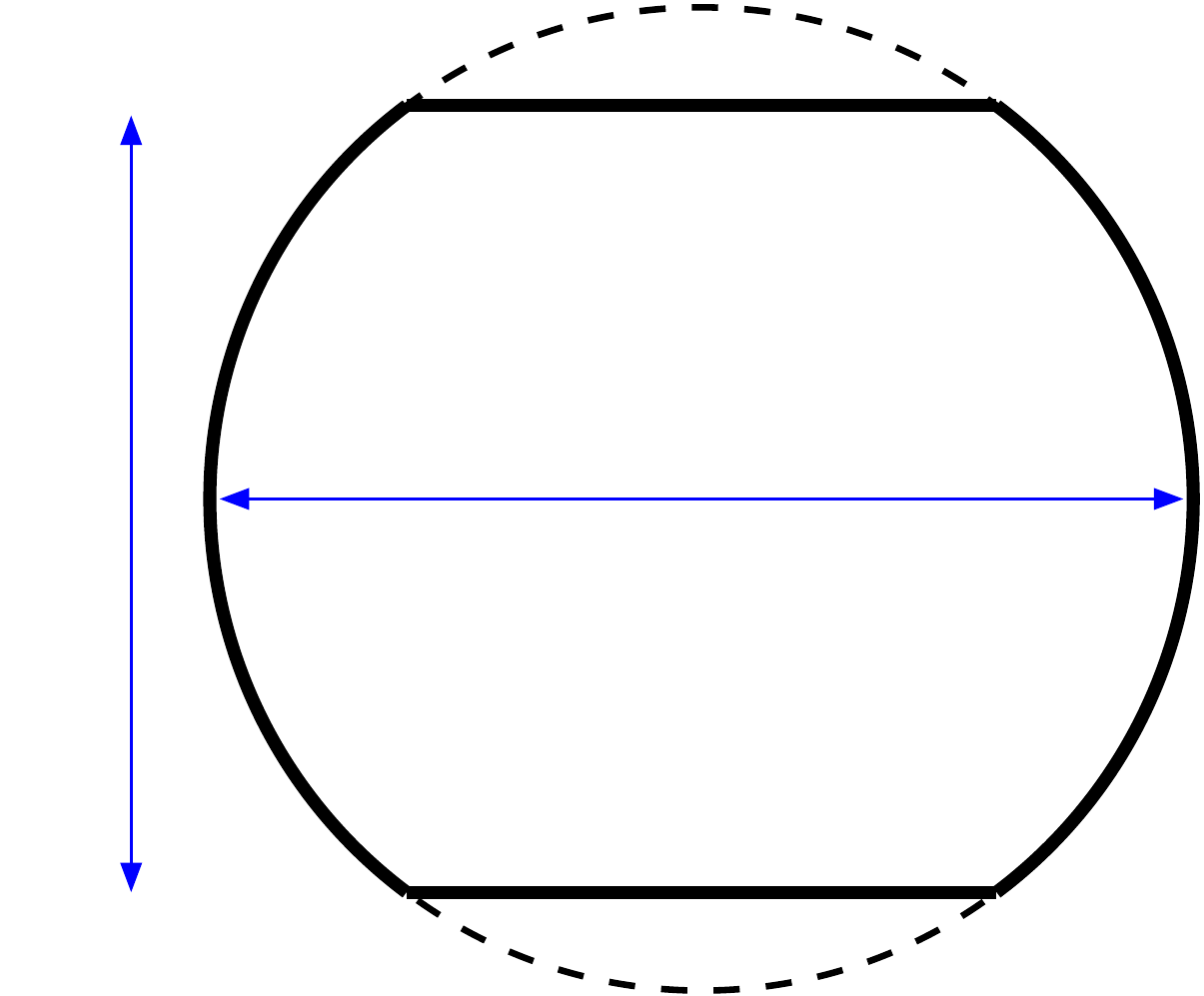}
}
\caption{A cut-sphere is a geometrical object obtained by chopping off  two symmetrical caps from a hard sphere of diameter $\Sigma$  at a distance (thickness) $L\leqslant\Sigma$.
 } \label{fig:cutspheres}
\end{figure}

To elucidate the effects of the particle shape upon the magnitude of entropic interaction, we simulate two cut-spheres in a bath of smaller hard spheres of diameter $\sigma$ (the depletant). A cut-sphere (CS) is a geometrical object obtained by chopping off two symmetrical caps from a hard sphere of diameter $\Sigma$  at a distance (thickness) $L\leqslant\Sigma$ (figure~\ref{fig:cutspheres}). {We use $\sigma$ as the unit length and choose $L/\sigma=8$ and $\Sigma/\sigma=10$.}

To study the system, we employ Monte Carlo simulations for hard bodies.
Since hard bodies have either a null interaction when they are separate
or an infinite repulsion when they overlap, in a hard-body Monte Carlo all
the moves are accepted unless they lead to the overlapping of particles.
Hence, to implement the simulation code, we need three overlap-checking procedures:
one to detect the overlaps among two hard spheres (interaction among the depletant particles),
one to detect the overlaps among a hard sphere and a CS (depletant-CS interaction)
and one to detect the overlaps among two CSs~\cite{Veerman92,Blaak99}.

We perform Monte Carlo simulations at a depletant packing fraction $\phi=0.00$, 0.05, 0.10, 0.15 with $10^{4}$ depletant
spheres for the highest $\phi$. {For each $\phi$, we perform from $10$ to $20$ independent simulations to reduce
the errors in the estimates for our observables.}

In each Monte Carlo step, we perform one attempt of {random displacement} for each depletant particle, $10$ attempts of {random displacements} for each CS and $10$  attempts of {random rotations} for each CS. {To perform random rotations, we employ the algorithm of} \cite{ATbook1989} {for linear molecules.} The attempted displacement and rotation for the CSs are tuned to reach an acceptance ratio of $\sim40\%$ during thermalization phase.

To ensure thermalization for simulated systems,  we check for two conditions to apply: (i) the mean square displacement $\left\langle \Delta r^{2}\right\rangle$ of the CS's is of the order of the simulation box length squared and (ii) the distribution of $\cos\theta_{\alpha}$ is flat, where $\theta_{\alpha}$ is the angle between the axis of a CS and the $\alpha$-axis ($\alpha=x,y,z$).

After thermalising the system, we sample the observables over $10^{11}$ Monte Carlo steps per run; within such time-scales, conditions (i) and (ii) are largely verified, hence ensuring a complete exploration of the configuration space for the two CSs.

Notice that our simulations involve only two CSs at a time; hence, we explore only two-body interactions of CSs' systems. Such an approach corresponds to the investigation of the infinite dilution limit for CSs (not for the depletant!), where three-body and higher interactions among CSs can be disregarded.

\section{Results}

We first study the radial distribution function $g(r)$ of the center to center distance $r$ between two CSs
\begin{equation}
g(r) = \rho^{-1} \frac{p(r)}{ \, 4 \pi r^2}\,,
\end{equation}
where $p(r)$ is the probability that the two CSs are at a distance $r$ and $\rho$ is the number density. To calculate $g(r)$, we evaluate via histograms $p(r)$ and then apply the discretised version of identity $p(r)\rd r = \rho \, g(r) \, 4\pi r^2\rd r$.

\begin{figure}[!b]
\centerline{
\includegraphics[width=0.64\textwidth]{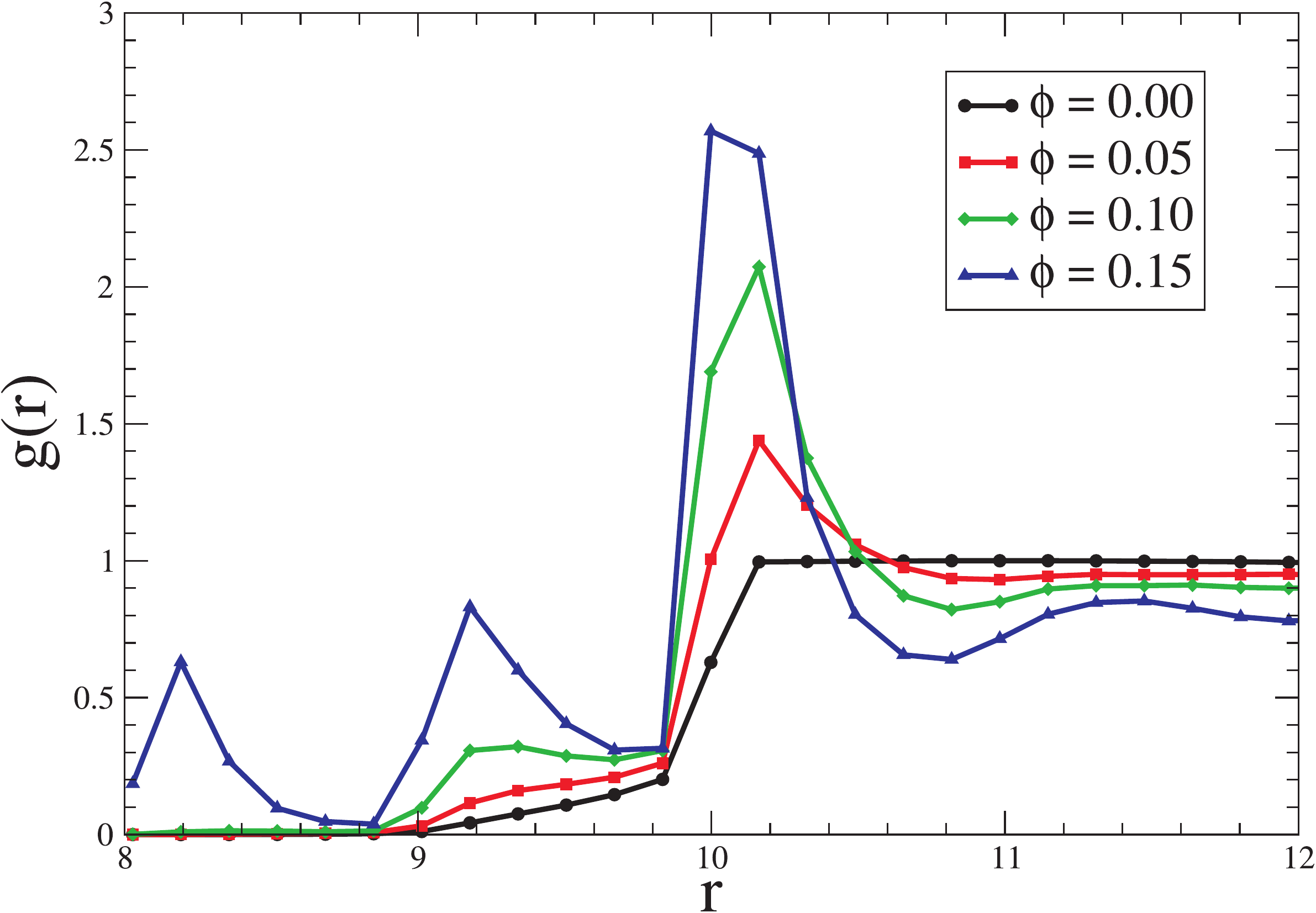}
}
\caption{(Color online) Radial distribution function $g(r)$ for the center-to-center distance $r$ between two CSs at different values of the packing fraction $\phi$. For $\phi=0$, only geometrical constrains matter (i.e., CSs cannot overlap); hence, $g(r)$ decreases when the two CSs approach the minimum possible distance $r=L$. For increasing $\phi$, depletion interactions make CSs attract and peaks develop in the region $L<r<\Sigma$.
} \label{fig:gofr}
\end{figure}

Figure~\ref{fig:gofr} shows that increasing $\phi$, the depletion interaction attracts CSs nearby. In fact, we notice that increasing $\phi$, the radial distribution function develops peaks at $ r \sim 10$ (i.e., $ r \sim  \Sigma$), at $r \sim 9/2$ (i.e. $ r \sim (L+\Sigma)/2$) and at $ r \sim 8$ (i.e. $ r \sim  L$).

By first-principle calculations of the work to separate two particles at an infinite distance, it can be shown \cite{ChandlerBook87} that the radial distribution function is related to the two-particle potential of the mean force $V_\textrm{eff}$ by $g(r)=\exp\left[-\beta\, V_\textrm{eff}\left(r\right)\right]$; such potential $V_\textrm{eff}$ accounts for integrating out all the other degrees of freedom (particles). In our case, the effective interaction among the CSs is  purely due to entropic effects since we are considering hard-bodies.
{A simple way of estimating the potential among two particles in the dilute limit is to calculate the work $W$ to bring such particles from distance $r$ to infinity: $g(r)=\exp\left[-\beta\, W\left(r\right)\right]$}\citep{ChandlerBook87}.
{Hence, to evaluate the magnitude of the entropic force while explicitly factoring out  the $r>L$ hard-body constrain, we calculate $V_\textrm{eff}$ from the formula:}
\begin{equation}
g(r)=g^\textrm{HS}_0\left(r\right)\,\exp\left[-\beta\, V_\textrm{eff}\left(r\right)\right]\, ,
\end{equation}
{where to avoid the divergent contribution of the hard-core, we have factored out the radial distribution function of hard spheres of diameter $L$ in the infinite dilute limit $g^\textrm{HS}_0$.}

\begin{figure}[!t]
\centerline{
\includegraphics[width=0.64\textwidth]{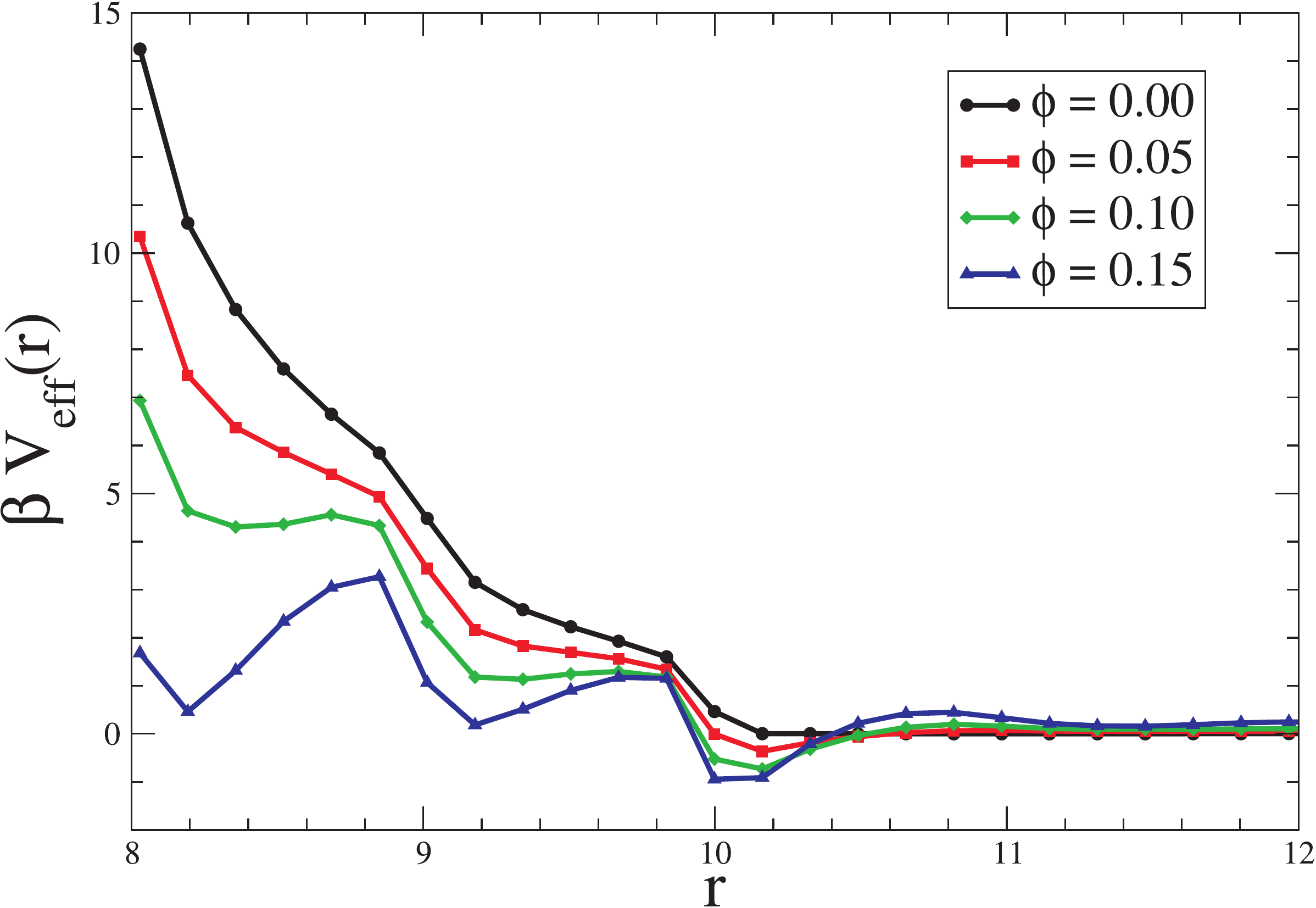}
}
\caption{(Color online)  Effective interaction $V_\textrm{eff}(r)$ among two CSs at different values
of the packing fraction $\phi$. For $\phi=0$, only geometrical constrains matter and $V_\textrm{eff}(r)$
increases when the two CSs approach the minimum possible distance $r=L$; {in fact, the number of possible
non-overlapping orientations~--- and hence the entropy~--- of two CSs decreases with a decreasing distance.}
Attractive depletion interactions  develop local minima at $ r \sim \Sigma$, at $r \sim (L+\Sigma)/2$
and at $ r \sim L$ for increasing $\phi$'s.
} \label{fig:betaVphi}
\end{figure}

In figure~\ref{fig:betaVphi} we show $V_\textrm{eff}$ among two CSs at different packing fractions. Notice that the evolution of $V_\textrm{eff}$  with $\phi$ shows competition among two terms: the attractive depletion interaction due to small spheres and the repulsive entropic interaction due to the geometry of the two CSs.
In particular, at $\phi=0$, no depletion interaction is present and $V_\textrm{eff}$ reflects the entropic contribution due to the fact that when two CSs are at a distance $r < \Sigma$ not all the orientations are possible; in particular, the number of possible orientations vanishes for $r=L$ where the $\phi=0$ potential has a repulsive maximum.

\begin{figure}[!b]
\centerline{
\includegraphics[width=0.64\textwidth]{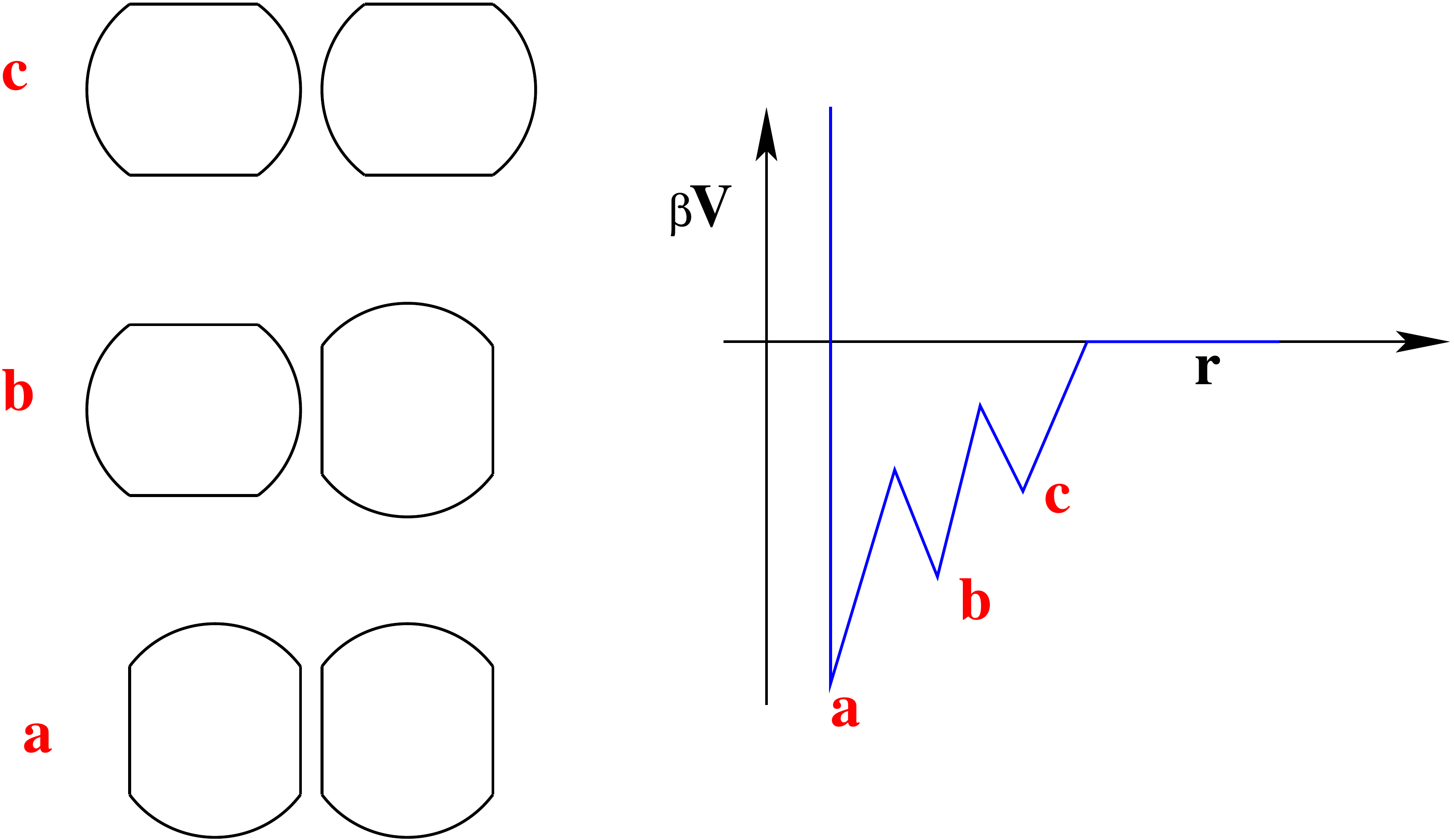}
}
\caption{(Color online)
Sketch of {the attractive part of the effective potential} induced by  a depletion interaction: since the attraction is proportional to the overlap volume, for a couple of cut-spheres there is maximum when flat faces are at contact (case (a), distance $\sim L$), minimum when curved faces are at contact (case (c), distance $\sim\Sigma$), intermediate when a flat face is at contact with a curved face (case (b), distance $\sim(L+\Sigma)/2$).
Such configurations are observed when conducting visual inspection of the CSs configurations during the Monte Carlo runs {at non-zero $\phi$'s}.
} \label{fig:attractionSketch}
\end{figure}

On the other hand, when the two CSs approach, the depletion interaction is influenced by the local curvature of the surfaces of the particles coming at contact: in particular, it is maximum when flat surfaces come together and minimum when curved surfaces come together. Hence, for high enough $\phi$, the depletion interaction wins over the entropic repulsion and CSs have an overall attractive potential; we sketch in figure~\ref{fig:attractionSketch} the behaviour of the depletion potential based on these geometric arguments and confirmed by visual inspection of the CSs configurations during the Monte Carlo runs.

\section{Discussion}

We have investigated, via Monte Carlo simulations, the depletion interaction among cut-spheres. Cut-spheres are a simple model system of shaped particles with only two curvatures. We show that even in this case, a complex, directional potential arises from the geometry of a system, hence indicating the possibility of controlling depletion-driven assembly by engineering the shape of the particles, and, in particular, the local curvature of their surfaces.

By varying the number of flat regions (cuts) in the spheres, interactions can be made selective and directional, hence implementing a variant of the so-called patchy colloidal particles. In patchy particles, inter-particle interactions depend on the specific regions of the particles that come close to contact, leading to assembled structures with substantially fewer defects than via conventional self-assembly methods \cite{GlotzerPatchy04,Young2013} and enriching both the equilibrium phase diagrams and the possible out-of-equilibrium behaviours of the system \cite{Bianchi2006}.
Finally, at a low packing fraction, the presence of short-range repulsive potentials makes the system akin to core-softened potentials systems \cite{BuldyrevPhysA2002}, whose phase behaviour is definitely unusual for one-component systems, since they can show iso-structural solid-solid and/or liquid-liquid phase transitions \cite{ScalaJStatPhys2000}.

Our simulation approach is in the Asakura-Oosawa spirit and permits to capture the phenomenology of the shape induced depletion interaction; nevertheless, an important advance would be to develop analytical approaches similar to the ones already introduced for the depletion interaction of spherical particles, such as  density functional theory \cite{Gotzelmann1998}, virial expansions \cite{Mao1995}, the Derjaguin approximation \cite{Henderson2002}, the integral equation formulation of liquid state theory \cite{Henderson1986} or Laplace transforms of the radial distribution functions \cite{Trokhymchuk2001}.
In particular, it would be important to take into account the oscillatory behaviour due to excluded volume correlation effects following the entropic attraction arising at short distances and predicted by accurate numerical \cite{Dickman1997,Dijkstra1999} and theoretical \cite{Attard1990,Gotzelmann1998} calculations of the effective potential between hard-sphere colloids and depletants.

\subsection*{Acknowledgements}

AS thanks D. Blaak for his help in setting up and testing the algorithm for checking the overlap among cut-spheres.


\ukrainianpart

 \title{Спрямовані збіднювальні взаємодії у сформованих частинках}
 \author{А. Скаля\refaddr{label1},
  П.Г. Де Санктіс Лучентіні\refaddr{label2}}
 \addresses{
 \addr{label1} Фізичний факультет, Університет ``Ла Сап'єнца'', 00185 Рим, Італія
 \addr{label2} Російський державний Університет нафти і газу ім. Губкіна, Москва, РФ}

 \makeukrtitle

 \begin{abstract}
 \tolerance=3000%
 Сили ентропії в колоїдних суспензіях і в системах полімер-колоїд привертають постійну увагу вже впродовж багатьох років.
 Експерименти показують як ентропійні сили можна використати для контролю самоскупчення колоїдних частинок.
 Значний прогрес, досягнутий в області синтезу колоїдних частинок впродовж минулих двох десятиріч, уможливлює виготовлення наночастинок
 високої якості з добре контрольованими розмірами, формами і складами і свідчить про те, що такі частинки можуть бути використані як
 ``штучні атоми'' для створення нових матеріалів.
 Для того, щоб пояснити вплив форми частинок на величину ентропійної взаємодії, ми аналізуємо ентропійну взаємодію двох розрізаних сфер.
 Ми показуємо, що розчинник індукує спрямоване збіднювальне притягання між плоскими поверхнями розрізаних сфер. Такий вплив
 вказує на можливість використання форми частинок для контролю спрямованості і сили взаємодії.
 \keywords ентропійні взаємодії, самоскупчення, Монте Карло
 \end{abstract}

 \end{document}